\documentstyle[12pt]{article}

\def\fun#1#2{\lower3.6pt\vbox{\baselineskip0pt\lineskip.9pt
        \ialign{$\mathsurround=0pt#1\hfill##\hfil$\crcr#2\crcr\sim\crcr}}}

\newcommand\eq[1]{Eq.~(\ref{#1})}
\newcommand\eqs[2]{Eqs.~(\ref{#1}) and (\ref{#2})}

\newcommand\ee{\end{equation}}
\newcommand\be{\begin{equation}}
\newcommand\eea{\end{eqnarray}}
\newcommand\bea{\begin{eqnarray}}

\newcommand\mm{\,\mbox{mm}}
\newcommand\cm{\,\mbox{cm}}

\newcommand\TeV{\,\mbox{TeV}}
\newcommand\GeV{\,\mbox{GeV}}

\newcommand\mpl{M_{\rm P}}

\newcommand\lsim{\mathrel{\rlap{\lower4pt\hbox{\hskip1pt$\sim$}}
    \raise1pt\hbox{$<$}}}
\newcommand\gsim{\mathrel{\rlap{\lower4pt\hbox{\hskip1pt$\sim$}}
    \raise1pt\hbox{$>$}}}

\def\dslash{\not{\hbox{\kern-2pt $\partial$}}}
\def\Dslash{\not{\hbox{\kern-4pt $D$}}}
\def\Oslash{\not{\hbox{\kern-4pt $O$}}}
\def\Qslash{\not{\hbox{\kern-4pt $Q$}}}
\def\pslash{\not{\hbox{\kern-2.3pt $p$}}}
\def\kslash{\not{\hbox{\kern-2.3pt $k$}}}
\def\qslash{\not{\hbox{\kern-2.3pt $q$}}}

 \newtoks\slashfraction
 \slashfraction={.13}
 \def\slash#1{\setbox0\hbox{$ #1 $}
 \setbox0\hbox to \the\slashfraction\wd0{\hss \box0}/\box0 }
 
\def\ee{\end{equation}}
\def\be{\begin{equation}}

\newcommand\sub[1]{_{\rm #1}}

\begin{document}

\begin{flushright}
LANCS-TH/9821
\\hep-ph/9810320\\
(October 1998)
\end{flushright}
\begin{center}
{\Large \bf Inflation with TeV-scale gravity \\
needs supersymmetry}

\vspace{.3in}
{\large\bf  David H.~Lyth}

\vspace{.4 cm}
{\em Department of Physics,\\
Lancaster University,\\
Lancaster LA1 4YB.~~~U.~K.}

\vspace{.4cm}
{\tt E-mail: d.lyth@lancaster.ac.uk 
}
\end{center}

\vspace{.6cm}
\begin{abstract}

Allowing for the possibility of 
large extra dimensions, the
fundamental Planck scale $M$ could be anywhere in the range
$\TeV\lsim M\lsim \mpl$, where 
$\mpl=2.4\times 10^{18}\GeV$ is the four-dimensional Planck scale.
If $M\sim\TeV$, quantum corrections would not destabilize the Higgs mass
even if there were no supersymmetry. But we point out that
supersymmetry must in fact be present, if there is an era of 
cosmological inflation, since during such an era the
inflaton mass satisfies
$m\ll M^2/\mpl=10^{-15}(M/\TeV)$ and supersymmetry will be needed to 
protect it.
If the inflation hypothesis is accepted, there is no
reason to think that 
Nature has chosen the low value $M\sim \TeV$, however convenient that
choice might have been for the next generation of collider experiments.

\end{abstract}

1. There is a large hierarchy, $m\sub H/\mpl\sim 10^{-15}$, between
the mass $m\sub H\sim 1\TeV$ of the Standard Model Higgs field and 
the Planck scale $\mpl\equiv (8\pi G)^{-1/2} \simeq 10^{18}\GeV$.
This makes it difficult to understand the existence of the Higgs field,
because in a generic field theory valid 
up to the Planck scale every elementary scalar field will have
a mass of order $\mpl$. To be precise, the mass will be given by
$m^2=m_0^2 + \sum\Delta_i +
\cdots$, where the first term is the tree-level value and the
$\Delta_i$ are one-loop contributions of order
$(\lambda_i\mpl)^2$, with $\lambda_i$ the strength of the interaction.
To avoid this disaster, it is usual to invoke supersymmetry which 
automatically cancels  the one-loop contributions to 
sufficient accuracy.\footnote
{The alternatives are an accidental cancellation, perhaps to be 
understood anthropically \cite{barr}, or a composite Higgs such as 
occurs in technicolor theories. It is difficult to construct composite 
theories that are consistent with observation.}

An alternative proposal \cite{ADD} is to place the 
fundamental Planck scale $M$ in the TeV region.
This makes the 
one-loop contributions of order
$(\lambda_i M)^2$ and avoids fine-tuning. The low value of $M$ is 
achieved 
by invoking 
extra space dimensions with large compactification radius.
If there are $n$ extra space dimensions with compactification radius
$R$, and $M$ is  the fundamental Planck scale in $4+n$ spacetime dimensions,
the Planck scale that we observe is given by 
\be
\mpl^2\sim R^n M^{2+n} \,.
\label{fundplanck}
\ee
Einstein or Newtonian gravity holds on scales $\gsim R$, but is modified
on smaller scales. 
Putting $M\sim 1\TeV$ and $n\geq 2$ gives $R\gsim 1\mm$, which is 
allowed by observation since the law of gravity is unknown on scales
$\lsim 1\cm$. One can envisage more complicated schemes, where
the compactified dimensions have different radii, but in all cases
the biggest dimension must be $R\lsim 1\cm$.

In this scheme, the hierarchy $m\sub H/\mpl$ is replaced by the 
hierarchy $R^{-1}/M$, which is of a different type and might be
easier to understand. Investigations reported so far 
\cite{more} seem to suggest 
that the scheme is viable, with no need for supersymmetry 
in the field theory that contains the Standard Model.\footnote
{Other works, for example \cite{DDG}, have explored the possibility
of large extra dimensions within the context of supersymmetry.}
As I now explain, this apparent success disappears
as soon as one tries to construct a model of inflation, that is 
presumably necessary to generate structure in the Universe.
A more detailed investigation will be reported elsewhere
\cite{p98tev}.

2. We are concerned with the cosmology of the observable Universe.
The Universe is modeled as a practically homogeneous and isotropic fluid,
with the distance between comoving fluid elements proportional to a 
universal scale factor $a(t)$. The evolution of the scale factor is 
given by the Friedmann equation, which 
assuming spatial flatness is\footnote
{For simplicity, I am assuming that
$\mpl$ during inflation has its present value.
With $M$ declared fixed (the choice of energy unit) this amounts to 
saying that $R$ has its present value.
The more general possibility is considered elsewhere \cite{p98tev}.}
\be
3H^2 = \rho/\mpl^2 \,,
\label{fried}
\ee
and the continuity equation
\be
\dot\rho = - 3H (\rho + P) \,.
\ee
Here $H=\dot a/a$ is the Hubble parameter, $\rho$ is the energy 
density and $P$ is the pressure.

The Friedmann and continuity equations are consequences of 
Einstein's field equation, and are 
valid provided that all relevant quantities are smoothed on a 
comoving distance scale $\gg R$. The
cosmic fluid, which is the 
subject of cosmology, is defined at a given epoch only after such
smoothing. (There is no question of `modifying the
Friedmann equation on short distance scales' since we are dealing with
a  universal scale factor.)

The history of the Universe begins at some energy
density $\lsim M^4$, where
 $M\sim \TeV$ is the fundamental Planck scale.
In particular, the potential $V\simeq \rho $ during inflation satisfies
\be
V(\phi)\lsim M^4 \,.
\label{vbound}
\ee
The vacuum fluctuation of the inflaton 
field $\phi$, that is supposed to be the origin of large scale
structure, is generated on each comoving scale $a/k$ at the epoch of 
horizon exit $k=aH$. We therefore require 
the Hubble distance $H^{-1}$ to be much bigger than the radius of the 
internal dimensions, $HR\ll 1$. Because of \eq{vbound}, this
is not a very severe restriction. 
Since $M^{-1}$ is the smallest distance that makes sense in the 
context of quantum gravity, we must have $RM\gsim 1$.
Then \eqs{vbound}{fried} 
give $H^{-1}\gsim 1\mm$,
whereas observation requires $R\lsim 1\cm$.
Irrespective of observation, \eqs{fundplanck}{vbound}
require $(RH)^2\lsim 1$ if $n\geq 2$
\cite{p97toni}. (Indeed, they give $(RH)^2\sim (MR)^{2-n}$.)
 The inclusion of additional, smaller 
dimensions only strengthens this result.

If $m $ is the inflaton mass 
during inflation,\footnote
{During inflation, the `vacuum' for quantum field theory is defined by 
the values of the inflaton and other relevant fields, which may be 
different from their true vacuum values. In hybrid inflation models
\cite{hybrid}
a field $\psi$ has a coupling like $\psi^2\phi^2$, which holds it at
the origin during inflation. Afterwards, $\psi$ acquires its true vacuum 
value, which can give the inflaton a large mass in the vacuum.
As a result, the quantum field theory containing Standard Model particles and 
the inflaton could have all scalar masses of order $\TeV$
{\em in the true vacuum}.
But we need a sensible quantum field theory also during 
inflation.}
the potential is of the form 
\be
V=V_0 \pm \frac12m^2\phi^2 + \cdots \,.
\label{vpot}
\ee
The dots represent additional terms, which might 
come from a variety
of sources \cite{p97toni} (higher powers of $\phi$ representing 
interaction terms in the tree-level potential, logarithmic terms 
representing loop corrections etc.). 
In order to generate the nearly scale-invariant 
primordial curvature perturbation, that is
presumed to be the origin of large scale structure, 
one should have \cite{p97toni}
\be
\left\vert \frac{\mpl^2 V''}{V} \right\vert \ll 1 \,,
\label{flat}
\ee
while cosmological scales are leaving the horizon.

Since each term has a different $\phi$ dependence, and $\phi$
usually varies significantly over cosmological scales, there is hardly 
likely to be an accurate cancellation between terms. 
Then, discounting the possibility $\phi\gg\mpl$ which would surely
place the field theory out of control, the flatness condition has
to be satisfied by each term individually, with $V$ dominated by the 
constant term $V_0$. In particular, the mass has to satisfy
$m^2\ll V_0/\mpl^2$. Remembering that $V_0\lsim M^4$ this becomes \cite{bd}
$m^2\ll M^4/\mpl^2 \sim (1\mm)^{-2}$, or
\be
\frac{m}{M} \ll \frac{M}{\mpl} \sim 10^{-15 } \,.
\ee
 
To summarise, taking the fundamental Planck scale $M$ to be of order
$1\TeV$ removes the hierarchy between the Higgs mass and $M$, at
the expense of introducing at least the {\em same} hierarchy
between the inflaton mass and $M$. To protect the 
inflaton mass from quantum corrections, supersymmetry is needed,
just as it was needed to protect the Higgs mass in the case
$M=\mpl$. 

In order to reheat the Universe, the inflaton must have 
significant couplings (not necessarily tree-level) with 
Standard Model particles. As a result, these particles should belong to 
the same supersymmetric field theory as the inflaton.\footnote
{By contrast, the Kaluza-Klein tower of scalar particles, associated
with the extra dimensions, need not be considered as part of the same
theory since they will have very weak coupling. For the same reason, 
their masses are presumably not destabilized by loop corrections.}

If one accepts the hypothesis of cosmological inflation, the
original motivation \cite{ADD} 
for considering  $M\sim 1\TeV$ is now removed,
and there seems to be no reason why Nature should have chosen this value.
Still, one may choose to explore that possibility,
either
because  it will be accessible to observation in the forseeable future
or because a lot of effort has been invested in it. 

3. 
In that case, one might ask whether a viable model of inflation
can be constructed.
It is easy enough to write down a potential
$V(\phi)\lsim M^4$, valid during slow-roll inflation, 
that gives the correct curvature perturbation $\delta_H
=1.9\times 10^{-5}$ on COBE scales and a spectral index
within the observed band $|n-1|<0.2$. Take, for instance, the potential
\be
V=V_0-\frac14\lambda \phi^4 + \cdots\,,
\ee
with the additional terms negligible during slow-roll inflation.
Let us assume 
that 
\be
\phi\sub{end}/\phi\sub{COBE}\gg 1 \,,
\label{ratio}
\ee
where $\phi\sub{end}$ is the end of slow-roll inflation and $\phi\sub{
COBE}$ is the epoch when COBE scales leave the horizon.
Using well-known formulas \cite{p97toni}, the
COBE constraint is $\lambda
\sim 10^{-14}(30/N)^2$ independently of $V_0$, and 
$n-1=-3/N$. Here $N$ is the number
of $e$-folds after COBE scales leave the horizon, given
(discounting thermal inflation and late-decaying particles)
by 
\be
N \simeq 30 - \ln(1\TeV/V^{1/4}) - \frac13\ln(V^{1/4}/T\sub{reh})  
\,.
\ee
Moreover, $\phi\sub{COBE} \simeq 10^{-5} \GeV$, so the initial
assumption \eq{ratio} is not very restrictive.

One might also wish to impose the constraint $\phi\lsim M\sim \TeV$;
for instance this might be necessary to have control over 
non-renormalizable terms if they are of order $\phi^d/M^{d-4}$,
or to have control over the  running of couplings and masses
in a renormalizable theory. In the above example 
this constraint no problem, but in general it is a severe restriction
as will be discussed elsewhere \cite{p98tev}.

4. It is generally accepted that a viable cosmology should begin with an
era of inflation, to set suitable initial conditions for the subsequent 
hot big bang and in particular to provide an origin for structure.
We have argued that the inflaton mass during inflation has to satisfy
$m\ll (M^2/\mpl) = 10^{-15}(M/\TeV)$, and that supersymmetry 
should be invoked to stabilize this
mass (or the masses of scalar fields produced after inflation).
A more detailed investigation \cite{p98tev} supports this conclusion.
If one accepts it, along with the need for inflation, one concludes that
there is no reason to think that
Nature has chosen $M\sim\TeV$, however convenient such a choice might 
have been for the next generation of collider experiments.

\section*{Acknowledgements}
This work was initiated at CERN. I
am grateful to CERN for support, and 
to Steve Abel, Karim Benakli, Keith Dienes, 
Tony Gerghetta, 
John March-Russell and Toni Riotto for useful discussions and correspondence.

\def\pl#1#2#3{{\it Phys. Lett. }{\bf B#1~}(19#2)~#3}
\def\zp#1#2#3{{\it Z. Phys. }{\bf C#1~}(19#2)~#3}
\def\prl#1#2#3{{\it Phys. Rev. Lett. }{\bf #1~}(19#2)~#3}
\def\rmp#1#2#3{{\it Rev. Mod. Phys. }{\bf #1~}(19#2)~#3}
\def\prep#1#2#3{{\it Phys. Rep. }{\bf #1~}(19#2)~#3}
\def\pr#1#2#3{{\it Phys. Rev. }{\bf D#1~}(19#2)~#3}
\def\np#1#2#3{{\it Nucl. Phys. }{\bf B#1~}(19#2)~#3}
\def\mpl#1#2#3{{\it Mod. Phys. Lett. }{\bf #1~}(19#2)~#3}
\def\arnps#1#2#3{{\it Annu. Rev. Nucl. Part. Sci. }{\bf #1~}(19#2)~#3}
\def\sjnp#1#2#3{{\it Sov. J. Nucl. Phys. }{\bf #1~}(19#2)~#3}
\def\jetp#1#2#3{{\it JETP Lett. }{\bf #1~}(19#2)~#3}
\def\app#1#2#3{{\it Acta Phys. Polon. }{\bf #1~}(19#2)~#3}
\def\rnc#1#2#3{{\it Riv. Nuovo Cim. }{\bf #1~}(19#2)~#3}
\def\ap#1#2#3{{\it Ann. Phys. }{\bf #1~}(19#2)~#3}
\def\ptp#1#2#3{{\it Prog. Theor. Phys. }{\bf #1~}(19#2)~#3}

\end{document}